\documentclass{PoS}
\usepackage{amsmath}

\title{Charmonium spectrum including higher spin and exotic states}

\ShortTitle{Charmonium spectrum including higher spin and exotic states}

\author{\speaker{C.~Ehmann}, G.~Bali \\
        Institut f\"ur Theoretische Physik, Universit\"at Regensburg, D-93040 Regensburg, Germany.\\
        E-mail: \email{christian.ehmann@physik.uni-regensburg.de}}

\abstract{We study the charmonium spectrum including higher spin
and exotic states. We use the Sheikholeslami-Wilson (clover) action
for $N_f=2$ sea quarks as well as for the charm
valence quark. In order to access excited
states we apply
a variational method with a basis of highly optimized operators.}

\FullConference{The XXV International Symposium on Lattice Field Theory\\
		 July 30-4 August 2007\\
		 Regensburg, Germany}

\begin{document}

\section{Introduction}
The discovery of about ten new charmonium resonances within the past
five years prompted an increased experimental and theoretical
interest in the phenomenology of these states. This trend is bound to
continue, in particular since the PANDA experiment at 
the new antiproton facility FAIR will produce huge new
data samples whose interpretation will require theoretical input.
One way both to reproduce and to predict the experimental spectrum of
particle resonances from first principles QCD are Lattice simulations.

Properties of many recently observed states are at variance with
nonrelativistic quark model predictions, the most striking example
being the X(3872).
It is an important task to reveal the inner structure of such states,
i.e.\ to clarify to what extent
charmonium resonances can be interpreted as
quark model $c\bar{c}$ states or whether some of these contain
significant quark-gluon hybrid or
four quark contributions (tetraquark or molecule).

We expect the lightest
hybrid charmonium states to be heavier than 4.3 GeV,
the experimental $D\overline{D}$ threshold lies above 3.7~GeV
and the vector charmonium $J/\psi$ ground state much lower,
at about 3.1~GeV.
This means that, with the exception of $D$
waves and higher angular momentum states, 
the higher Fock components will only start to show up prominently in
radial excitations. As a first step, we need to be able to reliably
compute these on the lattice.

We describe the methods used to obtain these states, our simulation
set-up and present first results on the
charmonium spectrum and ``wavefunctions''.

\section{Variational method}
We start from 
a cross correlator matrix 
\begin{equation}
C_{ij}(t) = \langle O_i(t) {O}_j^{\dagger}(0) \rangle,
\end{equation}
with a basis of operators $O_i$, $i=1,\ldots,N$,
destroying a colour singlet state within the desired
lattice $O_h\otimes C$ representation
from which we wish to deduce 
continuum quantum numbers.
We do not include charmed sea quarks and hence
the numbers of charm and anti-charm quarks are
separately conserved. At present we restrict
ourselves to the $c\bar{c}$ sector (including hybrids).
At a later stage four quark operators will be
incorporated.

The correlator matrix can be spectrally decomposed,
\begin{equation}
 \label{specdec}
  C_{ij}(t) = \sum_n v_i^n v_j^{n*} e^{-E_nt},
\end{equation}
where $v^n$ is the $n$th state within the subsector of the
Hilbert space spanned by $C(t)$,
and $E_n$ is the corresponding energy eigenvalue.
Since $C_{ij}$ is a real
symmetric matrix\footnote{This only holds in the limit of infinite
statistics. We symmetrize $C(t)$ by hand, after
checking that violations are consistent with zero,
within the statistical errors.}, the $v^n$ are mutually orthogonal.

To obtain initial guesses of the eigenvalues $\lambda^{\alpha}$ and -vectors
$\psi^{\alpha}$, $\alpha =1,\ldots,N$,
we solve the generalized eigenvalue problem~\cite{Mi85,LuWo90},
\begin{equation}
\label{evp} 
 C^{-1/2}(t_0)C(t)C^{-1/2}(t_0)\psi^{\alpha}_{t_0}(t) = \lambda^{\alpha}_{t_0}(t)
\psi^{\alpha}_{t_0}(t)\,,
\end{equation}
varying $t_0$ and $t>t_0$. Due to this symmetrized construction the
$N$-component eigenvectors $\psi^{\alpha}$ are mutually orthogonal
for all choices of $t_0$ and $t$. Note that the eigenvalues of the
system
$C(t)\phi^{\alpha}=\lambda^{\alpha}C(t_0)\phi^{\alpha}$ with
$\psi^{\alpha}=C^{1/2}(t_0)\phi^{\alpha}$ are the same as in Eq.~(\ref{evp}). Moreover,
the non-orthogonal $\phi$s will approach the orthogonal
$\psi$s at large $t_0$.

If we choose $t_0$ too large, the rank of $C(t_0)$ will not
be maximal anymore as (within statistical errors) excited
states will die out in Euclidean time. For
$t_0$ chosen too small, $C(t)$ will receive contributions from
more than the $N$ lowest lying states, resulting in unstable
eigenvectors and effective masses,
\begin{equation}
m_{\mbox{\scriptsize eff}}^{\alpha}(t)=
a^{-1}\ln\left(\frac{\lambda(t)}{\lambda(t+a)}\right)\,,
\end{equation}
where we have suppressed the subscript $t_0$.
For each channel we employ a three dimensional basis of
operators which we call local, narrow and
wide.

We apply iterative Gaussian smearing to the fermion fields $\phi$,
\begin{equation}
\label{eq:gauss}
\phi^{(n+1)}_x=c\left(\phi^{(n)}_x+
\kappa\sum_{j=\pm 1}^{\pm 3}\overline{U}_{x,j}\phi^{(n)}_{x+a\hat{\boldsymbol{\jmath}}}\right)\,,
\end{equation}
with $\kappa=0.3$ and a normalisation $c$ to avoid
numerical overflow. We smear quark and antiquark with the same
number of steps $n_g$ which is equivalent to applying $2n_g$ smearing
steps to one propagator only.

\begin{figure}[!ht]
\begin{center}
\resizebox{300pt}{!}{\includegraphics[clip]{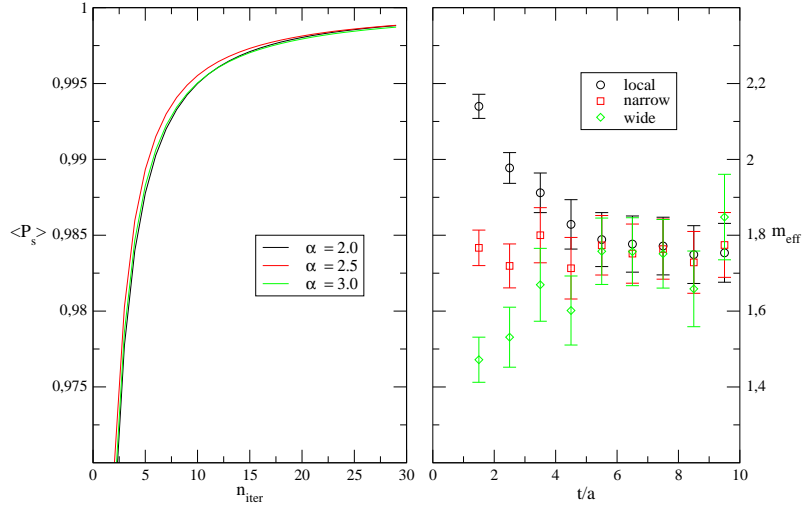}}
\end{center}
\caption{Dependence of the average value of the spatial plaquette
on the number of p-APE smearing steps (left) and
effective masses for local-smeared correlators with different
smearing functions (right).}
\label{apejacobi}
\end{figure}

The parallel transporters
$\overline{U}_{x,j}=U_{x,j}^{(15)}$  above are p-APE smeared:
\begin{equation}
\label{eq:smear}
U_{x,i}^{(n+1)}= P_{SU(3)}\left(U_{x,i}^{(n)}+\alpha\sum_{|j|\neq i}
U_{x,j}^{(n)}U^{(n)}_{x+a\hat{\boldsymbol{\jmath}},i}U^{(n)\dagger}_{x+a\hat{\boldsymbol{\imath}},j}\right)\,,
\end{equation}
where $\alpha=2.5$ was chosen to maximize the spatial plaquette
constructed from the smeared links
(see the left hand side of Figure \ref{apejacobi}).
$P_{SU(3)}$ denotes a projection
operator, back into the gauge group.
By using smeared transporters within Eq.~(\ref{eq:gauss})
we achieve a more continuum-like spatial distribution of the
smearing wavefunction (see Section~\ref{sec:wave} below) and
better overlaps with the physical states.

Effective masses for symmetric 2-point
functions with a local source and different sink smearings
are shown in Figure~\ref{apejacobi}.
We selected the smearing applied to the
trial wavefunctions within each channel,
such that one effective mass approached the asymptotic state
from above, one from below and ones sat exactly on spot.
This procedure ensures that the span of our variational basis
has overlap not only with the ground state but also with
the lowest radial excitations.

\vspace{0.5cm}

\section{Simulation details}
While chiral symmetry plays a minor role for charmonia,
the charm quark mass $m_c$ is not heavy enough to allow for a non-relativistic
treatment. We use the clover Wilson action both for valence
and sea quarks which will give us a well-defined continuum limit.
However, $m_ca\not\ll 1$, such that ultimately an $a\rightarrow 0$
extrapolation will be important.
We work on $N_f=2$ dynamical lattices generated by the QCDSF
collaboration. Details of these lattices can be found in
Ref.~\cite{Ali Khan:2003cu}. Here we present results
from a single $16^3\times 32$ lattice at $\beta=5.20$ and
$\kappa=0.13420$ corresponding
to an inverse lattice spacing of $a^{-1}\approx 1.73$ GeV
and a pion mass $m_{\pi}\approx 1$~GeV. The lattice spacing
was determined from the value $r_0\approx 0.46$~fm such
that the nucleon reaches its experimental mass when extrapolated
to physical $m_{\pi}$. This leaves us with the charm quark
mass as the only free parameter which we set
by tuning
$m_{\overline{1S}}=\frac{1}{4}m_{\eta_c}+\frac{3}{4}m_{J/\Psi}$
to the experimental value.

The operators we use are based on Ref.~\cite{Liao:2002rj},
however derivatives were symmetrized
to allow for charge conjugation eigenstates also at finite momenta.
The quark bilinears about which we report here
are displayed in Table \ref{operators}, together with
their irreducible lattice representations and
the lowest spin continuum state they couple to.

\begin{table}[h!]
\begin{center}
\begin{tabular}{|c|c|c|c|c|}
 \hline
 name & $O_h$ repr. & $J^{PC}$ & state & operator  \\
 \hline
 \hline
 $a_0$ & $A_1$ & $0^{++}$ & $\chi_{c0}$ & 1   \\
 $\pi$ & $A_1$ & $0^{-+}$ & $\eta_c$ & $\gamma_5$   \\
 $\rho$ & $T_1$ & $1^{--}$ & $J/\psi$ & $\gamma_i$  \\
 $a_1$ & $T_1$ & $1^{++}$ & $\chi_{c1}$ & $\gamma_5\gamma_i$  \\
 $b_1$ & $T_1$ & $1^{+-}$ & $h_c$ & $\gamma_i\gamma_j$  \\
 $(a_1\times\nabla)_{T_2}$ & $T_2$ & $2^{--}$ &  & $\gamma_5 s_{ijk}\gamma_j\nabla_k$  \\
 $(b_1\times\nabla)_{T_1}$ & $T_1$ & $1^{-+}$ & {\small exotic} & $\gamma_4\gamma_5\epsilon_{ijk}\gamma_j\nabla_k$  \\ 
\hline 
\end{tabular}
\end{center}
\caption{Interpolating fields in use.}
\label{operators}
\end{table}

\section{Spectrum}
\begin{figure}[ht!]
\begin{center}
\resizebox{300pt}{!}{\includegraphics[clip]{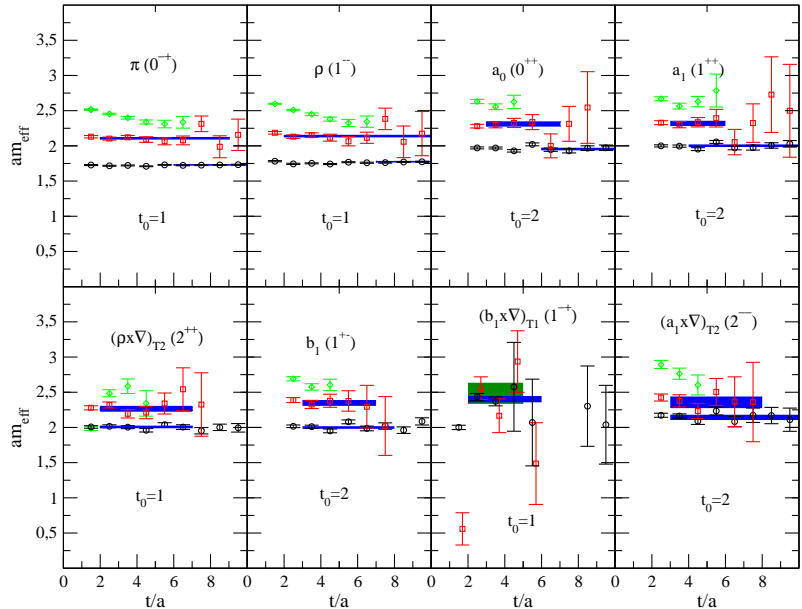}}
\end{center}
\caption{Effective masses from the three dimensional operator basis. Fit ranges
and errors are indicated by horizontal lines. The $t_0$ values
refer to the respective
normalization time slices~(see Eq.~(\protect\ref{evp})).}
\label{eff_mass}
\end{figure}

In Figure \ref{eff_mass} we show 
effective masses for the analyzed lattice.
In all channels we see nice plateaus for the ground and the first
excited state. Apart from the exotic $1^{-+}$ channel we also obtain reasonable
signals for the second excited state which however we digest with
caution: to gain more confidence in these we will
move to a larger operator basis and increase statistics.
The lines indicate the fit ranges and errors.
For the $1^{-+}$ it turned out particularly hard to separate the
ground state from the first excitation because these are very close
in mass. Hybrid potentials are rather flat and yield dense spectra
within potential models. Hence this maybe taken as a hint at
a hybrid content.
However, the creation operator with best overlap with this state
does not contain an explicit chromomagnetic field dependence: in a relativistic
theory, for any allowed $J$ all $PC$ quantum numbers including exotic ones
can be obtained from quark bilinears, even in the free field case.
More study of this question is required.

\begin{figure}[ht!]
\begin{center}
\resizebox{300pt}{!}{\includegraphics[clip]{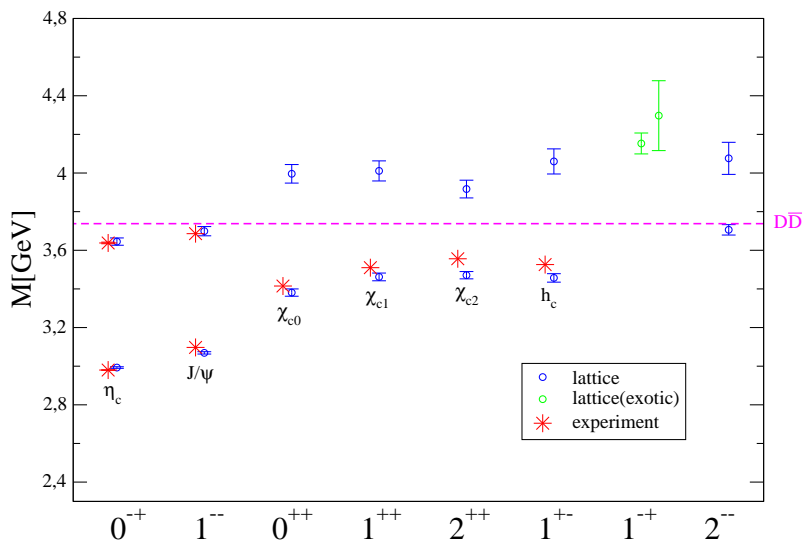}}
\end{center}
\caption{Predicted spectrum, together with the experimental values.
The $D\overline{D}$ threshold is the experimental one.}
\label{spectrum}
\end{figure}

The computed spectrum is
plotted in Fig.~\ref{spectrum},
together with the experimental values. It is important to note that
we were not too careful when setting the charm quark mass parameter
and underestimate $m_{\overline{1S}}$ by about 15~MeV.
However, this is still well within the accuracy of the
lattice spacing determination and moreover will cancel
from level splittings.
So the whole spectrum should be shifted higher a bit.

We obtain a 1S hyperfine splitting of $\Delta m_{1S}=73(2)$ MeV, below
the experimental $117$ MeV and consistent with the unphysically
high pion mass, wrong number of sea quarks and lattice artefacts.
Also disconnected contributions can affect this quantity.
For the 2S hyperfine splitting we obtain
$\Delta m_{2S}=47(6)$ MeV, in agreement with
experiment [49(5) MeV].
Our failure to consistently underestimate this
value as well might be explainable by $D\overline{D}$ threshold effects
which we neglect due to our heavy sea quarks (and hence $D$ mesons).
Alternatively, the $\eta_c$ mass might receive small contributions from
the $U_A(1)$ anomaly or from $\eta_c-\eta'$ mixing, also effects
that we neglect, whose full treatment requires disconnected quark line diagrams and charmed sea quarks.

\vspace{0.5cm}

\section{Wavefunctions}
\label{sec:wave}
The variational method not only helps to compute the spectrum but
also provides access to couplings. Efforts to extract such couplings
in the heavy quark regime have already been made for example in
Refs.\ \cite{Dudek:2007wv} and \cite{Burch:2007fj}. There exists no
real shortcut to the computation
of three-point functions for this purpose but
nonetheless it can be instructive to
analyze the smearing functions that we use in some detail.

\begin{figure}[h!]
\resizebox{140pt}{!}{\includegraphics[clip]{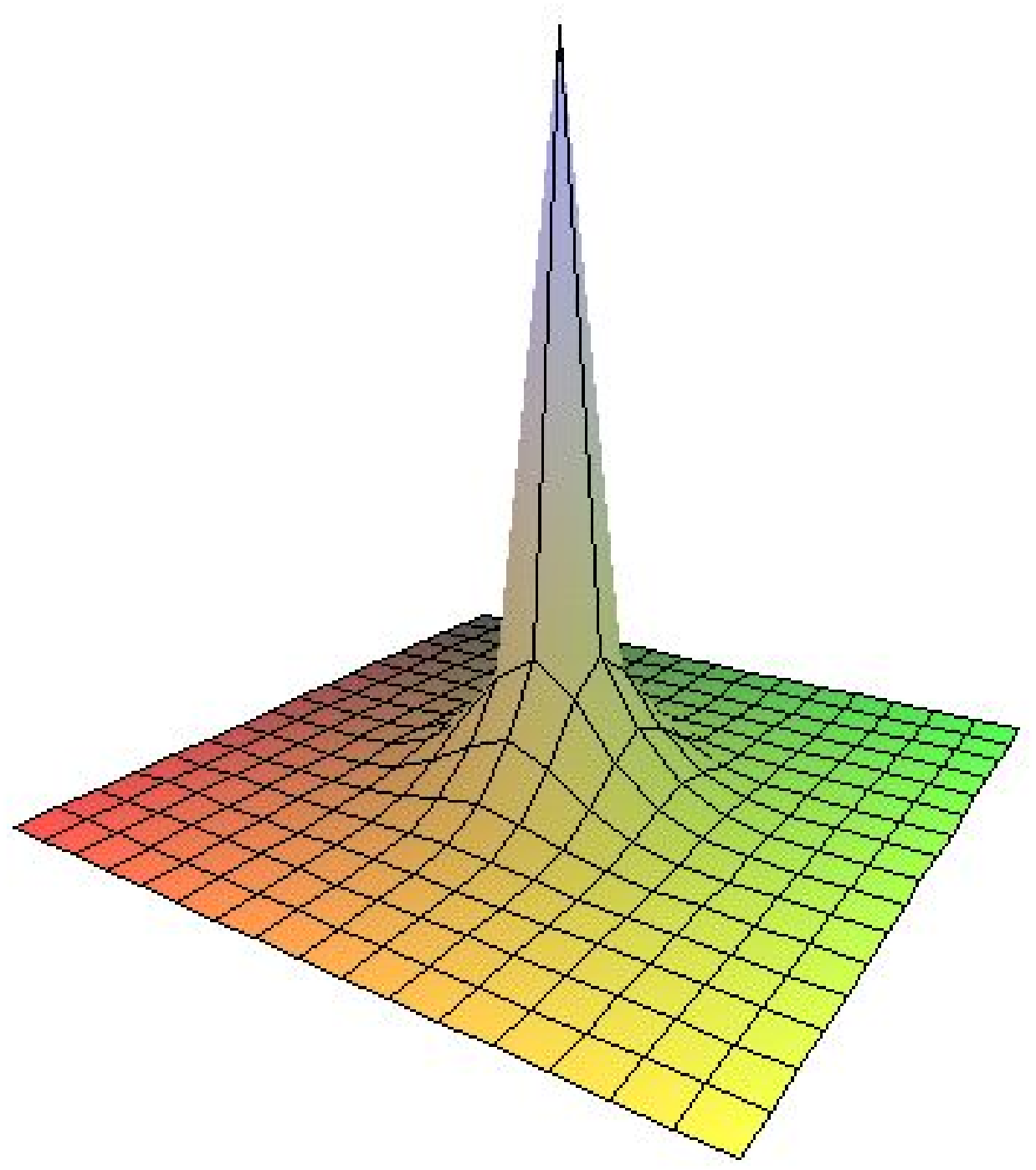}}
\resizebox{140pt}{!}{\includegraphics[clip]{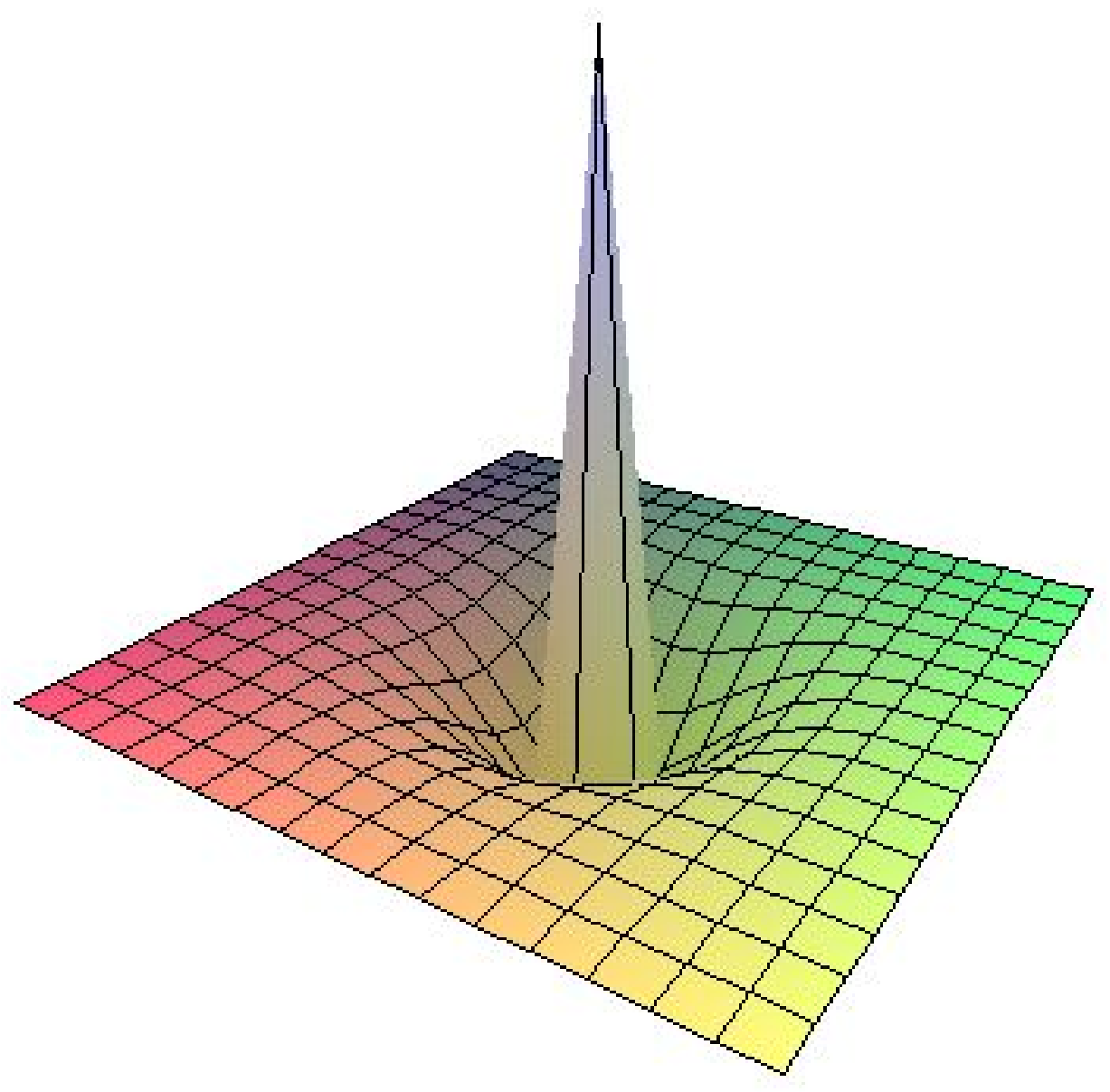}}
\resizebox{140pt}{!}{\includegraphics[clip]{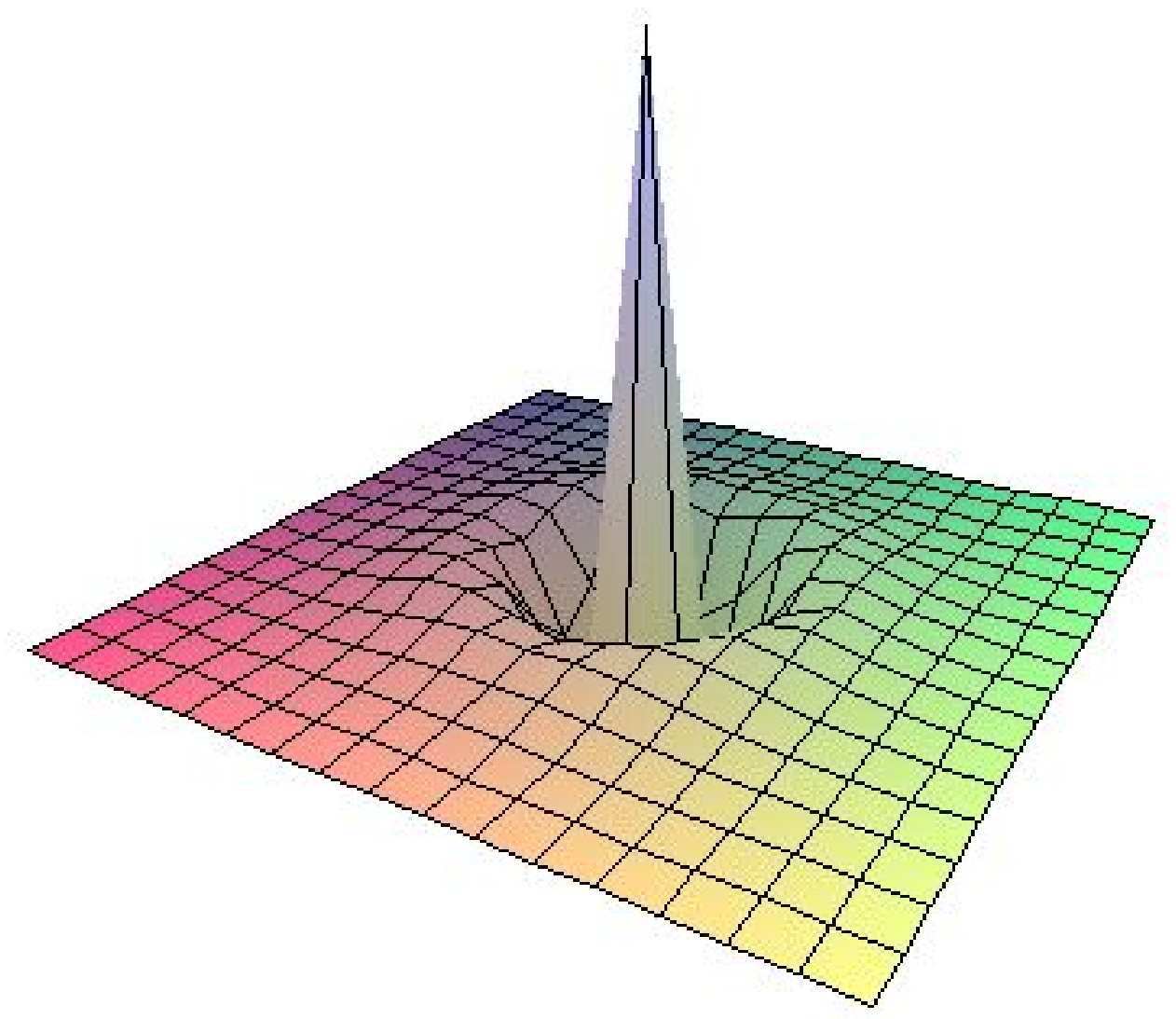}}
\caption{The 1S, 2S and 3S pseudoscalar "wavefunctions".}
\label{wf}
\end{figure}

For sufficiently large $t_0$ we can identify
the components of the eigenvectors $\psi^{\alpha}$ 
of Eq.~(\ref{evp}) as the couplings
of our interpolating functions with the physical state
of mass $m_{\alpha}$.
Here we start from a four-dimensional basis
of trial functions with 0, 5, 10 and 40 (times two) Gauss smearing iterations.
We can apply these functions to $\delta$ sources (in space and colour)
to obtain the spatial distribution of
the corresponding trial wavefunctions $\Phi_j({\mathbf x})\in SU(3)$.
By folding these with our eigenvectors
we can attempt to construct the ``wavefunctions" of the physical states:
$\Psi^{\alpha}({\mathbf x})=\sum_j\psi^{\alpha}_j\Phi_j({\mathbf x})$.
Needless to say that it is not possible to exactly
create the physical eigenstates
with such a small number of trial functions. 
Moreover, in Euclidean time we cannot obtain the phase information and hence
only the (gauge invariant) probability densities are meaningful
quantities. Thus in general $\mbox{Tr}\,\Psi^{\alpha\dag}\Psi^{\beta}\neq
\delta^{\alpha\beta}$.
Unfortunately, the used lattice is too coarse to resolve the node structure of
$|\Psi|^2$.
However, observing that $|\Psi|^2$ for our APE smeared fields and $|\Psi|^2$ in the free case are very similar,
we plot the wavefunctions for the free case, where the nodes are clearly visible due to the sign change.
Fig.~\ref{wf} shows these wavefunctions for the lowest three pseudoscalar states.
We neglect the statistical errors.

In spite of the small basis the node structure is consistent with
the $1S$, $2S$ and $3S$ assignment, with no visible pollution
from higher Fock states or $D$ waves. For the $1S$ we obtain an
{\em rms} width of ca.\ 0.39 fm. This compares reasonably
well with the infinite volume continuum potential model expectation of 
about 0.4 fm~\cite{Bali:1998pi}.

\vspace{1cm}

\section{Outlook}
The next step is to include disconnected contributions to the
charmonium two-point functions which may play an important role for
quantities like the $1S$ hyperfine splitting. Work on improved
all-to-all propagator calculations
is already in progress. We also plan to extend our operator
basis by including four-quark states, once we have moved to
light sea quark masses.

\acknowledgments
We thank the QCDSF Collaboration for making their
configurations available on the ILDG.
The runs were performed on the local QCDOC using the Chroma
software library (\cite{Edwards:2004sx, Boyle}).
This work is supported by EC Hadron Physics I3 Contract
RII3-CT-2004-506087, by BMBF Contract 06RY257
and by the GSI University Program Contract RSCHAE.

\end{document}